# Green functions from a gauge-invariant effective action for the electroweak Standard Model

A. Denner,[a] S. Dittmaier[b*] and G. Weiglein[a]

[a]Institut für Theoretische Physik, Universität Würzburg, Am Hubland, D-97074 Würzburg, Germany

[b]Theoretische Physik, Universität Bielefeld, Universitätsstraße, D-33501 Bielefeld, Germany

Application of the background-field method to the electroweak Standard-Model yields a gauge-invariant effective action giving rise to simple Ward identities. We find that in the background-field 't Hooft–Feynman gauge the resulting vertex functions are exactly those that are obtained using the pinch technique. Thus, the background-field method provides a general framework that generalizes the pinch technique directly and uniquely to arbitrary Green functions and arbitrary orders of perturbation theory. Moreover, the desirable properties of the pinch-technique vertex functions hold for arbitrary gauge parameters and have a simple explanation within the background-field method.

## 1. INTRODUCTION

All known successful theories describing the interactions of elementary particles are gauge theories. However, in order to evaluate quantized gauge theories within perturbation theory, one has to break gauge invariance in intermediate steps by choosing a definite gauge. As a consequence, although the physical observables, i.e. the S-matrix elements, are gauge-independent, the Green functions, the building blocks of the S-matrix elements, are gauge-dependent in the conventional formalism.

Before we proceed, we remind the reader of the notion of gauge invariance and gauge independence: gauge invariance means invariance under gauge transformations. The gauge invariance of the classical Lagrangian gives rise to Ward identities between the Green functions of the quantized theory. Gauge independence becomes relevant when quantization is done by fixing a gauge. It means independence of the method of gauge fixing.

The gauge dependence of Green functions poses no problem as long as one calculates physical observables in a fixed order of perturbation theory. However, as soon as one does not take into account all contributions in a given order one will in general arrive at gauge-dependent results. This happens usually if one tries to resum higher-order corrections via Dyson summation of self-energies or if one is only interested in particular contributions like definite formfactors, e.g. magnetic moments for off-shell particles, without taking into account the full set of diagrams. This has been frequently done in the literature.

Motivated by these facts, various attempts have been made to define gauge-independent building blocks. In order to construct gauge-independent running couplings, several proposals for gauge-independent self-energies have been put forward [1,2]. These were essentially obtained by considering four-fermion processes and shifting parts of the box and vertex diagrams to the self-energies to cancel the gauge-parameter dependence of the latter within the class of $R_\xi$ gauges. As one can shift arbitrary gauge-independent contributions between the different building blocks the resulting quantities are not unique. This freedom has been used to require certain desirable properties from the self-energies, like a decent asymptotic behaviour and the vanishing of the photon–Z-boson mixing at zero-momentum transfer. It nevertheless resulted in different definitions of gauge-independent building blocks. All these ad-hoc treatments only refer to four-fermion

*Supported by the Bundesministerium für Forschung und Technologie, Bonn, Germany.

processes and do not give a general prescription which is applicable to other vertex functions.

Such a prescription is given by the so-called pinch technique (PT) [3,4]. The PT is an algorithm for the construction of (within $R_\xi$ gauges) gauge-independent vertex functions by reorganizing parts of the Feynman diagrams contributing to a manifestly gauge-independent quantity, leaving only a trivial gauge dependence in the tree propagators. The results obtained via the PT directly fulfil the desirable properties that had to be explicitly enforced in the ad-hoc treatments mentioned above. But even more important, it turns out that the vertex functions constructed according to the PT fulfil the simple Ward identities related to the classical Lagrangian.

However, the PT leaves many questions unanswered. So far, it has only been realized for specific vertex functions at the one-loop level. Its application to other vertex functions is not always clear, and its generalization to higher orders is non-trivial and non-unique. Although the PT vertex functions are claimed to be process-independent this has to the best of our knowledge not been proven but only shown for specific examples. It is very unsatisfactory that no explanation exists for the fact that the PT rules yield vertex functions with desirable properties and in particular that these vertex functions fulfil simple Ward identities. Finally, although the application of the PT to four-fermion processes is rather simple, it turns out that the explicit construction of general PT vertex functions can be technically quite involved.

We pursue a different approach. We do not try to construct gauge-independent quantities by reorganizing Feynman graphs in the conventional formalism but we keep gauge invariance from the start. To this end we use the background-field method.

## 2. BACKGROUND-FIELD METHOD

The background-field method (BFM) [5] is a technique for quantizing gauge theories without losing explicit gauge invariance. In particular, it allows to construct a gauge-invariant effective action. This is done by decomposing the usual gauge field $V'$ in the classical Lagrangian into a quantum field $V$ and a background field $\hat{V}$

$$\mathcal{L}_{\text{class}}(V') \to \mathcal{L}_{\text{class}}(V + \hat{V}). \tag{1}$$

One adds a gauge-fixing term that breaks the gauge invariance of the quantum field but preserves the background-field gauge invariance

$$\mathcal{L}_{\text{GF}} = -\frac{1}{2\xi_Q}[D_\mu^{ab}(\hat{V})V^{b,\mu}]^2, \tag{2}$$

where $D_\mu^{ab}(\hat{V})$ is the covariant derivative with respect to the background field,

$$D_\mu^{ab}(\hat{V})V^{b,\mu} = \partial_\mu V^{b,\mu} + gf^{abc}\hat{V}_\mu^b V^{c,\mu}. \tag{3}$$

Here $g$ is the gauge coupling, and $f^{abc}$ are the structure constants of the gauge group. In this way one constructs an effective action $\Gamma[\hat{V}]$ which is invariant under gauge transformations of the background field,

$$\hat{V}_\mu^a \to \hat{V}_\mu^a - f^{abc}\omega^b\hat{V}_\mu^c + \frac{1}{g}\partial_\mu\omega^a, \tag{4}$$

where $\omega^a$ denote the (infinitesimal) parameters of the gauge transformation. This invariance gives rise to simple Ward identities between the vertex functions derived from $\Gamma[\hat{V}]$. The Ward identities can be directly obtained by evaluating

$$\frac{\delta\Gamma[\hat{V}]}{\delta\omega^a} = 0. \tag{5}$$

Note that these vertex functions depend on the quantum gauge parameter $\xi_Q$.

In the background field method, the S-matrix is constructed as usual by forming trees with vertices from $\Gamma[\hat{V}]$ connected by lowest-order background-field propagators. To define these propagators, one has to add a gauge-fixing term to $\Gamma[\hat{V}]$. This gauge-fixing term is not related to the term used to fix the gauge inside loop diagrams, and the associated gauge parameter $\hat\xi$ only enters tree propagators but not the higher-order contributions to the vertex functions. The equivalence of the S-matrix in the BFM to the conventional one has been proven in Ref. [6].

Calculations in the Glashow–Salam–Weinberg model are usually carried out in the 't Hooft gauge. This gauge-fixing term contains not only the gauge fields but also the Higgs fields and eliminates tree-level mixing between the gauge bosons and the corresponding unphysical Higgs bosons. In order to generalize this gauge to a form that respects background-field gauge invariance, also the Higgs field has to be split into a background and a quantum part. While the background field has the usual non-vanishing vacuum expectation value, the one of the quantum field is zero. Denoting the background fields with a caret, the background-field 't Hooft gauge-fixing term reads [7]

$$\mathcal{L}_{\text{GF}} = -\frac{1}{2\xi_Q^W}\left[(\partial_\mu \delta^{ac} + g_2 \varepsilon^{abc}\hat{W}_\mu^b)W^{c,\mu}\right.$$

$$\left. - ig_2\xi_Q^W \frac{1}{2}(\hat{\Phi}_i^\dagger \sigma_{ij}^a \Phi_j - \Phi_i^\dagger \sigma_{ij}^a \hat{\Phi}_j)\right]^2 \quad (6)$$

$$- \frac{1}{2\xi_Q^B}\left[\partial_\mu B^\mu + ig_1\xi_Q^B \frac{1}{2}(\hat{\Phi}_i^\dagger \Phi_i - \Phi_i^\dagger \hat{\Phi}_i)\right]^2,$$

where we have used the conventions of Ref. [8], and $\sigma^a$, $a = 1,2,3$, denote the Pauli matrices. This gauge-fixing term translates to the conventional one upon replacing the background Higgs field by its vacuum expectation value and omitting the background $SU(2)_W$ triplet field $\hat{W}_\mu^a$. Background-field gauge invariance restricts the number of quantum gauge parameters to two, $\xi_Q^W$ for $SU(2)_W$ and $\xi_Q^B$ for $U(1)_Y$.

The background-field vertex functions can be calculated using ordinary Feynman rules that distinguish between quantum fields and background fields. Whereas the quantum fields appear only inside loops, the background fields appear only in tree lines. Apart from the doubling of the gauge fields the background-field Feynman rules differ from the conventional ones only owing to the gauge-fixing and ghost terms, which affect only vertices that involve both background and quantum fields. As the gauge-fixing term is non-linear in the fields, the gauge-boson vertices become gauge-dependent. The fermion fields are treated as usual, they have the conventional Feynman rules, and there is no distinction between external and internal fields.

We have evaluated the complete Feynman rules within the background-field 't Hooft gauge (6) including the associated ghost terms for $\xi_Q = \xi_Q^W = \xi_Q^B$. Here we give only some typical examples. The coupling of one background Z boson to two quantum W bosons reads

$$\Gamma_{\mu\nu\rho}^{\hat{Z}W^+W^-,(0)}(k_1,k_2,k_3) = ie\frac{c_W}{s_W}$$
$$\times \left[g_{\nu\rho}(k_3-k_2)_\mu + g_{\mu\rho}\left(2k_{1\nu} + \left(1-\frac{1}{\xi_Q}\right)k_{2\nu}\right)\right.$$
$$\left. - g_{\mu\nu}\left(2k_{1\rho} + \left(1-\frac{1}{\xi_Q}\right)k_{2\rho}\right)\right], \quad (7)$$

the coupling between two background Z bosons and two quantum W bosons

$$\Gamma_{\mu\nu\rho\sigma}^{\hat{Z}\hat{Z}W^+W^-,(0)}(k_1,k_2,k_3,k_4) = -ie^2\frac{c_W^2}{s_W^2} \quad (8)$$
$$\left[2g_{\mu\nu}g_{\rho\sigma} - (g_{\mu\sigma}g_{\nu\rho} + g_{\mu\rho}g_{\nu\sigma})\left(1-\frac{1}{\xi_Q}\right)\right],$$

and the coupling of one background Z boson to two charged ghosts

$$\Gamma_\mu^{\hat{Z}\bar{u}^\pm u^\pm,(0)}(k_1,k_2,k_3) = \pm ie\frac{c_W}{s_W}(k_3-k_2)_\mu. \quad (9)$$

Despite the fact that, owing to the doubling of the fields, the Feynman rules seem to become more complicated, actual calculations become in fact simpler. This is in particular the case in the 't Hooft–Feynman gauge ($\xi_Q = 1$) where many of the vertices relevant for one-loop graphs simplify considerably [compare (7) and (8)]. While the number of diagrams contributing to a certain vertex function is approximately the same as in the conventional formalism, the diagrams themselves become easier to evaluate. Moreover, the number of diagrams contributing to the full (reducible) Green functions can be reduced by choosing an appropriate background gauge, e.g. the unitary gauge or a non-linear gauge [13].

## 3. EXPLICIT RESULTS

Based on the background-field Feynman rules, we have evaluated the gauge-boson self-energies, the fermion–gauge-boson vertices, and the triple gauge-boson vertices at one-loop order. For the special value $\xi_Q = 1$, we find that all our results coincide with those obtained in the PT [4,9,10]. For illustration, we give here the difference between the Z self-energy including tadpole contributions $T$ calculated within the BFM

for arbitrary $\xi_Q$ and the well-known conventional 't Hooft–Feynman-gauge result:

$$\left[\Sigma_T^{\hat{Z}\hat{Z}}(k^2) - \frac{eM_W}{s_W c_W^2 M_H^2}T\right]\bigg|_{BF}$$
$$- \left[\Sigma_T^{ZZ}(k^2) - \frac{eM_W}{s_W c_W^2 M_H^2}T\right]\bigg|_{conv,\ \xi=1} =$$
$$\frac{\alpha}{4\pi}(k^2 - M_Z^2)\frac{1}{12c_W^2 s_W^2 M_Z^4}\bigg[16M_W^4(1-\xi_Q)$$
$$+ 2\frac{M_W^2}{k^2}[7k^2 - M_Z^2 + \xi_Q(k^2 + M_Z^2)]$$
$$\times [A_0(\sqrt{\xi_Q}M_W) - A_0(M_W)]$$
$$- [2M_W^2(16M_W^2 - 10M_Z^2 - 8k^2)$$
$$- k^2(k^2 + M_Z^2)]B_0(k^2, M_W, M_W)$$
$$- [4M_W^2 + (k^2 + M_Z^2)](4M_W^2 \xi_Q - k^2)$$
$$\times B_0(k^2, \sqrt{\xi_Q}M_W, \sqrt{\xi_Q}M_W)$$
$$- 2[(1-\xi_Q)^2 M_W^4 + (10 - 2\xi_Q)M_W^2 k^2 + k^4]$$
$$\times \frac{k^2 + M_Z^2}{k^2}B_0(k^2, M_W, \sqrt{\xi_Q}M_W)\bigg] \quad (10)$$
$$\xrightarrow[\xi_Q \to 1]{} -\frac{\alpha}{\pi}(k^2 - M_Z^2)\frac{c_W^2}{s_W^2}B_0(k^2, M_W, M_W).$$

The last line is identical to the term obtained in the framework of the PT [4]. Note that (10) vanishes at $k^2 = M_Z^2$.

The manifest gauge invariance gives rise to simple Ward identities related to the classical Lagrangian, which hold in all orders of perturbation theory. Here we list some examples:

$$k^\mu \Gamma_{\mu\nu}^{\hat{A}\hat{A}}(k) = 0, \quad \text{i.e.} \quad \Sigma_L^{\hat{A}\hat{A}}(k^2) = 0, \quad (11)$$
$$k^\mu \Gamma_{\mu\nu}^{\hat{A}\hat{Z}}(k) = 0, \quad \text{i.e.} \quad \Sigma_L^{\hat{A}\hat{Z}}(k^2) = 0, \quad (12)$$
$$k^\mu \Gamma_\mu^{\hat{A}\hat{\chi}}(k) = 0, \quad \text{i.e.} \quad \Sigma^{\hat{A}\hat{\chi}}(k^2) = 0, \quad (13)$$
$$k^\mu \Gamma_{\mu\nu}^{\hat{Z}\hat{Z}}(k) - iM_Z \Gamma_\nu^{\hat{\chi}\hat{Z}}(k) = 0, \quad (14)$$
$$k^\mu \Gamma_\mu^{\hat{Z}\hat{\chi}}(k) - iM_Z \Gamma^{\hat{\chi}\hat{\chi}}(k) + \frac{ie}{2s_W c_W}\Gamma^{\hat{H}}(0) = 0, (15)$$
$$k^\mu \Gamma_{\mu\nu}^{\hat{W}^\pm \hat{W}^\mp}(k) \mp M_W \Gamma_\nu^{\hat{\phi}^\pm \hat{W}^\mp}(k) = 0, \quad (16)$$
$$k^\mu \Gamma_\mu^{\hat{W}^\pm \hat{\phi}^\mp}(k) \mp M_W \Gamma^{\hat{\phi}^\pm \hat{\phi}^\mp}(k) \pm \frac{e}{2s_W}\Gamma^{\hat{H}}(0) = 0,$$
$$(17)$$
$$k^\mu \Gamma_\mu^{\hat{A}\bar{f}f}(k,p,\bar{p}) = -eQ_f[\Gamma^{\bar{f}f}(\bar{p}) - \Gamma^{\bar{f}f}(-p)], \quad (18)$$

$$k^\mu \Gamma_\mu^{\hat{Z}\bar{f}f}(k,p,\bar{p}) - iM_Z \Gamma^{\hat{\chi}\bar{f}f}(k,p,\bar{p}) = \quad (19)$$
$$e[\Gamma^{\bar{f}f}(\bar{p})(v_f - a_f \gamma_5) - (v_f + a_f \gamma_5)\Gamma^{\bar{f}f}(-p)],$$
$$k^\mu \Gamma_\mu^{\hat{W}^\pm \bar{f}'f}(k,p,\bar{p}) \mp M_W \Gamma^{\hat{\phi}^\pm \bar{f}'f}(k,p,\bar{p}) =$$
$$\frac{e}{\sqrt{2}s_W}[\Gamma^{\bar{f}'f'}(\bar{p})\frac{1-\gamma_5}{2} - \frac{1+\gamma_5}{2}\Gamma^{\bar{f}f}(-p)], \quad (20)$$

$$k^\mu \Gamma_{\mu\rho\sigma}^{\hat{A}\hat{W}^+\hat{W}^-}(k,k_+,k_-) =$$
$$e[\Gamma_{\rho\sigma}^{\hat{W}^+\hat{W}^-}(k_+) - \Gamma_{\rho\sigma}^{\hat{W}^+\hat{W}^-}(-k_-)], \quad (21)$$
$$k_+^\rho \Gamma_{\mu\rho\sigma}^{\hat{A}\hat{W}^+\hat{W}^-}(k,k_+,k_-)$$
$$- M_W \Gamma_{\mu\sigma}^{\hat{A}\hat{\phi}^+\hat{W}^-}(k,k_+,k_-) = \quad (22)$$
$$e\left[\Gamma_{\mu\sigma}^{\hat{W}^+\hat{W}^-}(-k_-) - \Gamma_{\mu\sigma}^{\hat{A}\hat{A}}(k) + \frac{c_W}{s_W}\Gamma_{\mu\sigma}^{\hat{A}\hat{Z}}(k)\right],$$
$$k_-^\sigma \Gamma_{\mu\rho\sigma}^{\hat{A}\hat{W}^+\hat{W}^-}(k,k_+,k_-)$$
$$+ M_W \Gamma_{\mu\rho}^{\hat{A}\hat{W}^+\hat{\phi}^-}(k,k_+,k_-) = \quad (23)$$
$$-e\left[\Gamma_{\mu\rho}^{\hat{W}^-\hat{W}^+}(-k_+) - \Gamma_{\mu\rho}^{\hat{A}\hat{A}}(k) + \frac{c_W}{s_W}\Gamma_{\mu\rho}^{\hat{A}\hat{Z}}(k)\right].$$

We use the conventions of Ref. [8], all momenta and fields are incoming and for the 2-point functions only the momentum of the first field is given. Note that the vertex functions are one-particle irreducible, i.e. they contain no tadpole contributions; these appear explicitly as $\Gamma^{\hat{H}}(0)$. Within the PT these Ward identities have been partially verified [4,9,10].

The PT vertex functions have been found to possess very desirable properties. These hold for the BFM vertex functions not only for $\xi_Q = 1$ but for arbitrary $\xi_Q$. In contrast to the PT, these properties can be explained within the BFM by relating them to the Ward identities and/or other simple arguments like power counting. Some examples are given in the sequel.

The mere gauge invariance of the (physical) propagator poles guarantees that these are not shifted in the BFM (and thus in the PT) compared with the conventional formalism [see (10)]. The Ward identity (12) implies together with the analyticity of $\Sigma_{\mu\nu}^{\hat{A}\hat{Z}}(k)$ at $k^2 = 0$ the vanishing of the photon–Z-boson mixing at zero-momentum transfer, $\Sigma_T^{\hat{A}\hat{Z}}(0) = 0$, in analogy to $\Sigma_T^{\hat{A}\hat{A}}(0) = 0$. In contrast to the PT, the IR finiteness of the self-energies is obvious within the BFM.

The fermion–gauge-boson vertex functions including fermion wave-function renormalization are UV finite. This can be derived from (18), (19) and (20) together with power-counting arguments like in QED. As a consequence, the asymptotic behaviour for $|k^2|\to\infty$ of the running couplings defined directly via Dyson summation of the self-energies is automatically governed by the renormalization group [4]. This fact reflects the connection between the self-energy and vertex renormalization in the BFM [5].

We have confirmed all these properties by explicit computation of the relevant quantities for arbitrary finite values of $\xi_Q$. For illustration, we present the leading logarithmic terms of the transverse parts of the gauge-boson self-energies in the high-energy limit

$$\Sigma_T^{\hat A\hat A}(k^2) \underset{|k^2|\to\infty}{\sim} \frac{\alpha}{4\pi} 7 k^2 \log\frac{|k^2|}{\mu^2},$$

$$\Sigma_T^{\hat A\hat Z}(k^2) \underset{|k^2|\to\infty}{\sim} -\frac{\alpha}{4\pi}\frac{42 c_W^2+1}{6 c_W s_W} k^2 \log\frac{|k^2|}{\mu^2},$$

$$\Sigma_T^{\hat Z\hat Z}(k^2) \underset{|k^2|\to\infty}{\sim} \frac{\alpha}{4\pi}\frac{42 c_W^4+2c_W^2-1}{6 c_W^2 s_W^2} k^2 \log\frac{|k^2|}{\mu^2},$$

$$\Sigma_T^{\hat W\hat W}(k^2) \underset{|k^2|\to\infty}{\sim} \frac{\alpha}{4\pi}\frac{43}{6 s_W^2} k^2 \log\frac{|k^2|}{\mu^2}. \qquad (24)$$

Note that these are independent of $\xi_Q$.

The properties of the $At\bar t$ and $Zt\bar t$ vertices have been discussed within the PT in Ref. [9]. From our results for arbitrary $\xi_Q$ we have explicitly confirmed the validity of the Ward identities (18) and (19) for the top-quark and recovered the PT results for $\xi_Q = 1$. The BFM yields the same result as the conventional formalism for the magnetic-dipole-moment form factor (MDM) of the top quark. In contrast to the statements in Ref. [9] but in agreement with Ref. [11], we find that the MDM vanishes in the limit $|k^2|\to\infty$ for all $\xi_Q$. This has been checked both numerically and analytically. Moreover, it can be deduced for all renormalizable gauges by a simple power-counting argument.

In Ref. [10] the three-gauge-boson vertices were derived within the PT for W bosons coupled to conserved currents and the Ward identity (21) was verified for this case. These results cannot be used as suitable building blocks in processes where this restriction does not apply as e.g. in $\gamma\gamma \to W^+W^-$. Projecting our general BFM results to the case of conserved currents, we find agreement with (3.16) of Ref. [10].[1] For our general off-shell result we have explicitly checked the Ward identity (21) and the other two Ward identities (22) and (23) which have not been mentioned in the PT context. In the BFM the anomalous-magnetic-moment form factor $\Delta\kappa$ and the electric-quadrupole-moment form factor $\Delta Q$ as defined in Ref. [12] are IR-finite and have a decent high-energy behaviour, i.e. they vanish in the limit $|k^2|\to\infty$.

While the vanishing of $\Delta Q$ in the high-energy limit follows directly from power counting, the vanishing of $\Delta\kappa$ can be derived as follows: inserting the covariant decompositions of the vertex functions into the Ward identities (22) and (23) and considering the coefficients of the various tensor structures, we find that the $\Delta\kappa$ is directly related to a coefficient of the $AW\phi$ vertex and another coefficient of the $AWW$ vertex. As these coefficients correspond to tensors of dimension two and three, they are of dimension -1 and -2, respectively. According to power counting, these coefficients and therefore $\Delta\kappa$ have to vanish for $|k|^2 \to \infty$ in renormalizable gauges.

Whereas in Ref. [10] essentially only the AWW vertex was investigated, we have also considered the ZWW vertex. We found that it fulfils the same properties as the former, only the Ward identities involve in addition the $\chi WW$ vertex function.

## 4. CONCLUSION

We have shown that the BFM provides a systematic way — via direct application of Feynman rules — to obtain Green functions that are derived from a gauge-invariant effective action, fulfil simple Ward identities and, in comparison to their $R_\xi$-gauge counterparts, possess very desirable properties such as improved UV and IR properties and a decent high-energy behaviour.

By applying the BFM to those cases for which the PT has been used in the literature, we found that the PT results can be recovered within the BFM by putting the quantum gauge parameter $\xi_Q = 1$. Thus the BFM provides a simple systematic generalization of the PT to arbitrary Green functions and higher orders.

---

[1] This equation contains an incorrect relative sign. We confirmed this by checking the Ward identities and redoing the PT calculation.

Moreover, the calculation of the vertex functions is much simpler in the BFM than in the PT. While the rearrangement of different contributions in the PT is quite cumbersome and not clear for more complicated vertex functions, the calculation within the BFM is comparable or even simpler than the evaluation of the vertex functions in the conventional formalism. While the BFM Green functions are obviously process-independent, this fact has not been proven within the PT.

We have found in addition that all desirable properties of the vertex functions hold in the BFM for arbitrary values of $\xi_Q$. This means in particular that the choice $\xi_Q = 1$, corresponding to the PT, is not distinguished on physical grounds but only one of arbitrarily many equivalent possibilities. Of course, the background-field 't Hooft–Feynman gauge technically facilitates actual calculations.

Our results demonstrate that the requirement of gauge-parameter independence used in the PT and former treatments is not the criterion leading to well-behaved vertex functions. In fact, the background-field vertex functions depend on the additional gauge parameter $\xi_Q$. This ambiguity is as well present in the PT and all similar constructions, since the definite prescription to eliminate the gauge-parameter dependence is just a matter of convention. As a consequence, although the quantities constructed in the PT appear to be gauge-parameter-independent, they are not guaranteed to be physically meaningful.

On the other hand, we showed that the desirable properties of the background-field vertex functions are a direct consequence of the BFM Ward identities, which reflect the underlying gauge invariance. Therefore we propose to investigate the physical relevance of quantities like form factors or running couplings on the basis of these Ward identities.

## REFERENCES

1. D.C. Kennedy and B.W. Lynn, *Nucl. Phys.* **B322** (1989) 1;
   D.C. Kennedy et al., *Nucl. Phys.* **B321** (1989) 83;
   B.W. Lynn, Stanford University Report No. SU-ITP-867, 1989 (unpublished);
   D.C. Kennedy, in *Proc. of the 1991 Theoretical Advanced Study Institute in Elementary Particle Physics*, eds. R.K. Ellis et al. (World Scientific, Singapore, 1992), p. 163.
2. M. Kuroda, G. Moultaka and D. Schildknecht, *Nucl. Phys.* **B350** (1991) 25.
3. J.M. Cornwall, *Phys. Rev.* **D26** (1982) 1453 and in *Proc. of the French-American Seminar on Theoretical Aspects of Quantum Chromodynamics*, ed. J.W. Dash, (Centre de Physique Théorique, Report No. CPT-81/P-1345, Marseille, 1982);
   J.M. Cornwall and J. Papavassiliou, *Phys. Rev.* **D40** (1989) 3474;
   J. Papavassiliou, *Phys. Rev.* **D41** (1990) 3179.
4. G. Degrassi and A. Sirlin, *Phys. Rev.* **D46** (1992) 3104.
5. L.F. Abbott, *Nucl. Phys.* **B185** (1981) 189; *Acta Phys. Pol.* **B13** (1982) 33 and references therein.
6. L.F. Abbott, M.T. Grisaru and R.K. Schaefer, *Nucl. Phys.* **B229** (1983) 372.
7. G. Shore, *Ann. Phys.* **137** (1981) 262;
   M.B. Einhorn and J. Wudka, *Phys. Rev.* **D39** (1989) 2758.
8. A. Denner, *Fortschr. Phys.* **41** (1993) 307.
9. J. Papavassiliou and C. Parrinello, NYU-TH-93/05/02.
10. J. Papavassiliou and K. Philippides, *Phys. Rev.* **D48** (1993) 4255.
11. K. Fujikawa, B.W. Lee, A.I. Sanda, *Phys. Rev.* **D6** (1972) 2923.
12. E.N. Argyres et al., *Nucl. Phys.* **B391** (1993) 23.
13. M.B. Gavela et al., *Nucl. Phys.* **B193** (1981) 257.